\documentclass[twocolumn,tighten]{aastex62}
\pdfoutput=1 

\shorttitle{\textit{NICER} Timing of PSR J1412+7922 and PSR J1849$-$0001}
\shortauthors{Bogdanov et al.}
\submitjournal{The Astrophysical Journal}

\begin{document}

\title{\textit{NEUTRON STAR INTERIOR COMPOSITION EXPLORER} X-RAY TIMING OF THE \\ RADIO AND $\gamma$-RAY QUIET PULSARS PSR J1412+7922 AND PSR J1849$-$0001}

\correspondingauthor{Slavko Bogdanov}
\email{slavko@astro.columbia.edu}

\author{Slavko Bogdanov}
\affiliation{Columbia Astrophysics Laboratory, Columbia University, 550 West 120th Street, New York, NY, 10027, USA}

\author{Wynn C.~G.~Ho}
\affiliation{Department of Physics and Astronomy, Haverford College, 370 Lancaster Avenue, Haverford, PA 19041, USA}
\affiliation{Mathematical Sciences, Physics and Astronomy, and STAG Research Centre, University of Southampton, Southampton SO17 1BJ, UK}

\author{Teruaki Enoto}
\affiliation{The Hakubi Center for Advanced Research and Department of Astronomy, Kyoto University, Kyoto 606-8302, Japan}

\author{Sebastien Guillot}
\affiliation{IRAP, CNRS, 9 avenue du Colonel Roche, BP 44346, F-31028 Toulouse Cedex 4, France}
\affiliation{Universit\'{e} de Toulouse, CNES, UPS-OMP, F-31028 Toulouse, France}

\author{Alice K.~Harding}
\affiliation{Astrophysics Science Division, NASA Goddard Space Flight Center, Greenbelt, MD 20771, USA}

\author{Gaurava K.~Jaisawal}
\affiliation{National Space Institute, Technical University of Denmark,  Elektrovej 327-328, DK-2800 Lyngby, Denmark}

\author{Christian Malacaria}
\affiliation{NASA Marshall Space Flight Center, NSSTC, 320 Sparkman Drive, Huntsville, AL 35805, USA}\thanks{NASA Postdoctoral Fellow}
\affiliation{Universities Space Research Association, NSSTC, 320 Sparkman Drive, Huntsville, AL 35805, USA}

\author{Sridhar S.~Manthripragada}
\affiliation{Instrument Systems and Technology Division, NASA Goddard Space Flight Center, Greenbelt, MD 20771, USA}

\author{Zaven Arzoumanian}
\affiliation{X-Ray Astrophysics Laboratory, NASA Goddard Space Flight Center, Greenbelt, MD 20771, USA}

\author{Keith C.~Gendreau} 
\affiliation{X-Ray Astrophysics Laboratory, NASA Goddard Space Flight Center, Greenbelt, MD 20771, USA}

\begin{abstract}
We present new timing and spectral analyses of PSR J1412$+$7922 (Calvera) and PSR J1849$-$0001, which are only seen as pulsars in X-rays, based on observations conducted with the \textit{Neutron Star Interior Composition Explorer} (\textit{NICER}). We obtain updated and substantially improved pulse ephemerides compared to previous X-ray studies, as well as spectra that can be well-fit by simple blackbodies and/or a power law. Our refined timing measurements enable deeper searches for pulsations at other wavelengths and sensitive targeted searches by LIGO/Virgo for continuous gravitational waves from these neutron stars.  Using the sensitivity of LIGO's first observing run, we estimate constraints that a gravitational wave search of these pulsars would obtain on the size of their mass deformation and r-mode fluid oscillation. 

\end{abstract}

\keywords{pulsars: general --- pulsars: individual (PSR J1412$+$7922, PSR J1849$-$0001) --- stars: neutron --- X-rays: stars --- gravitational waves}

\section{Introduction} \label{sec:intro}
The vast majority of the $\sim$2600 known rotation-powered pulsars have been discovered and have had their spin parameters determined using observations at radio wavelengths \citep{2005AJ....129.1993M}. However, an appreciable subset of pulsars in the Galaxy are observationally inaccessible in the radio due to severe dispersion/scattering or unfavorable viewing geometry such that the radio emission beams never intersect our line of sight. In addition, it is possible that some pulsars are intrinsically radio quiet. Regardless of the reason, sensitive observations for such objects at other wavelengths such as X-rays provide the only means of discovering them and characterizing their spin behavior.

The high Galactic latitude ($b=+37^{\circ}$) X-ray source 1RXS J141256.0$+$792204 (also known as Calvera or PSR J1412$+$7922) was identified by \citet{2008ApJ...672.1137R} and \citet{2009ApJ...705..391S} as a strong neutron star candidate based on its soft X-ray spectrum and lack of optical counterpart. It was revealed with \textit{XMM-Newton} to be a relatively nearby $P = 59.2$ ms pulsar  with nearly sinusoidal pulsations  \citep{2011MNRAS.410.2428Z}. The X-ray emission from Calvera can be described by either a purely thermal spectrum or a composite thermal plus non-thermal model. \citet{2013ApJ...778..120H} subsequently determined its period derivative to be $\dot{P}=(3.19 \pm 0.08)\times 10^{-15}$ s s$^{-1}$, which corresponds to characteristic age $\tau_c \equiv P/2\dot{P}=2.9\times 10^5$ yr, spin-down luminosity $\dot{E}\approx 6\times 10^{35}$ erg s$^{-1}$,  and surface dipole magnetic field strength $B_s = 4.4 \times 10^{11}$ G.  While these inferred properties are not unusual for a rotation-powered pulsar, radio searches have failed to detect pulsations down to fairly deep limits \citep[$>$0.05 mJy; see ][]{2007A&A...476..331H,2011MNRAS.410.2428Z}. PSR J1412$+$7922 does not appear to be a $\gamma$-ray source either, despite its likely proximity ($D\lesssim 2$ kpc) and relatively high spin-down luminosity  \citep[][]{2011ApJ...736L...3H,2013ApJ...778..120H}. Based on these characteristics, this pulsar is speculated to be a possible descendant of the Central Compact Object (CCO) class, a population of enigmatic radio-quiet young neutron stars in supernova remnants \citep[see, e.g.,][]{2013ApJ...765...58G,2017JPhCS.932a2006D}.
Most recently, using \textit{Chandra} HRC, \citet{2015ApJ...812...61H} measured a proper motion for PSR J1412$+$7922 of $\mu=69\pm26$ mas yr$^{-1}$, with a direction away from the Galactic plane.

The 38.5 ms X-ray pulsar PSR J1849$-$0001 was discovered in a targeted \textit{Rossi X-ray Timing Explorer} (\textit{RXTE}) observation of the soft $\gamma$-ray/TeV source IGR J18490$-$0000/HESS J1849$-$000 \citep{2011ApJ...729L..16G}.
The measured spin-down rate $\dot{P}=1.42\times 10^{-14}$ s s$^{-1}$ implies that PSR J1849$-$0001 is quite an energetic ($\dot{E}=9.8\times10^{36}$ ergs s$^{-1}$) and young ($\tau_c=42.9$ kyr) rotation-powered pulsar \citep{2011ApJ...729L..16G,2015MNRAS.449.3827K}.
The source exhibits a non-sinusoidal single pulse per rotation with width $\sim 0.5$ in rotation phase at both low energies (0.06--10~keV) and high energies (2--28~keV) and pulsed fractions of $0.77\pm0.04$ and $0.88\pm0.08$, respectively (also $0.76\pm0.02$ in the 2--10~keV band; \citealt{2015MNRAS.449.3827K}).
The hard non-thermal spectrum from a 23~ks 2012 \textit{Chandra} ACIS-S observation can be fit with absorption $N_{\rm H}=(4.30\pm0.16)\times10^{22}\mbox{ cm$^{-2}$}$ and power law index $\Gamma=1.08\pm 0.02$. A 54~ks 2011 \textit{XMM-Newton} spectrum can be fit with $N_{\rm H}=(4.5\pm0.1)\times10^{22}\mbox{ cm$^{-2}$}$ and $\Gamma\approx 1.2$ and yields a 2--10~keV unabsorbed flux $F_{2-10}^{\rm unabs}\approx4.8\times10^{-12}$ ergs cm$^{-2}$ s$^{-1}$, which implies X-ray luminosity $L_X=2.8\times10^{34}$ ergs s$^{-1}$ for an assumed distance of $D=7$ kpc (\citealt{2015MNRAS.449.3827K}; see also \citealt{2018arXiv180204833V}, who analyzed the same data and obtain similar results).
The ACIS-S data and a 25~ks 2011 HRC-S observation do not show evidence of extended emission from a pulsar wind nebula (PWN) at soft energies (0.1--2 keV and 0.5--2 keV, respectively) around PSR~J1849$-$0001, but the ACIS-S data shows diffuse (2--10 keV) emission $\lesssim 30^{\prime\prime}$ in extent and several 0.5--10~keV point sources $\sim 1^{\prime}$ from the pulsar. The 2011 \textit{XMM-Newton} observation shows much fainter diffuse emission  $75^{\prime\prime}$--$150^{\prime\prime}$ from PSR~J1849$-$0001; this emission contributes 13\% of the total flux around the pulsar and has a spectrum that can be fit by a power law with $\Gamma=1.75\pm0.05$ \citep{2015MNRAS.449.3827K}; note that \citet{2018arXiv180204833V} find a larger diffuse contribution (23\%), but this is likely due to different regions considered ($40^{\prime\prime}$--$100^{\prime\prime}$) and $N_{\rm H}$.
Finally, we note that \textit{Fermi} LAT does not detect pulsed emission from PSR~J1849$-$0001 \citep{2013ApJS..208...17A}.

These two pulsars are of additional interest as potential sources of continuous gravitational waves (GWs) that may be detectable by the Laser Interferometer Gravitational-Wave Observatory (LIGO) and Virgo. Neutron stars can be sources of continuous GWs at 2 or $\approx 4/3$ times the pulsar spin frequency, depending on the GW emission mechanism, and stars with significant asymmetry and faster spin are stronger GW emitters (see, e.g., \citealt{2017arXiv170907049G,2017MPLA...3230035R}). Previous GW searches of known pulsars used timing information from primarily radio and $\gamma$-ray observations \citep{2014ApJ...785..119A,2017ApJ...839...12A,2017PhRvD..96l2006A}.  Because PSR~J1412$+$7922 and PSR~J1849$-$0001 are only seen to be pulsed in X-rays, the timing models presented here will enable LIGO/Virgo to search for GWs from these two pulsars.

Herein we present X-ray studies of PSR J1412$+$7922 (Calvera) and PSR J1849$-$0001 using \textit{NICER}. The paper is organized as follows.
In Section~\ref{sec:observations}, we present details of the observations and data reduction procedures. We present the X-ray timing analysis in Section~\ref{sec:timing} and spectral analysis in Section~\ref{sec:spectra}. We provide conclusions in Section~\ref{sec:conclusions}.

\section{Observations} \label{sec:observations}

\textit{NICER} observations of a given target are typically carried out in segments lasting hundreds to $\sim$2000 seconds. All exposures of a target during the same UTC day are grouped into a single Observation ID (ObsID). The set of \textit{NICER} observations of PSR J1412$+$7922 and PSR J1849$-$0001 analyzed here are summarized in Tables A1 and A2 of the Appendix. The \textit{NICER} observations of these targets were often opportunistic, filling gaps in the schedule around higher-priority targets, such as transients. For PSR J1412$+$7922, the resulting data span 380 days from 2017 September 15 to 2018 October 3, while for PSR J1849$-$0001 the data span 223 days from 2018 February 13 to 2018 September 29.

\begin{figure}
\centering
\includegraphics[clip, trim=1.0cm 11cm 0.0cm 3.0cm,width=0.48\textwidth]{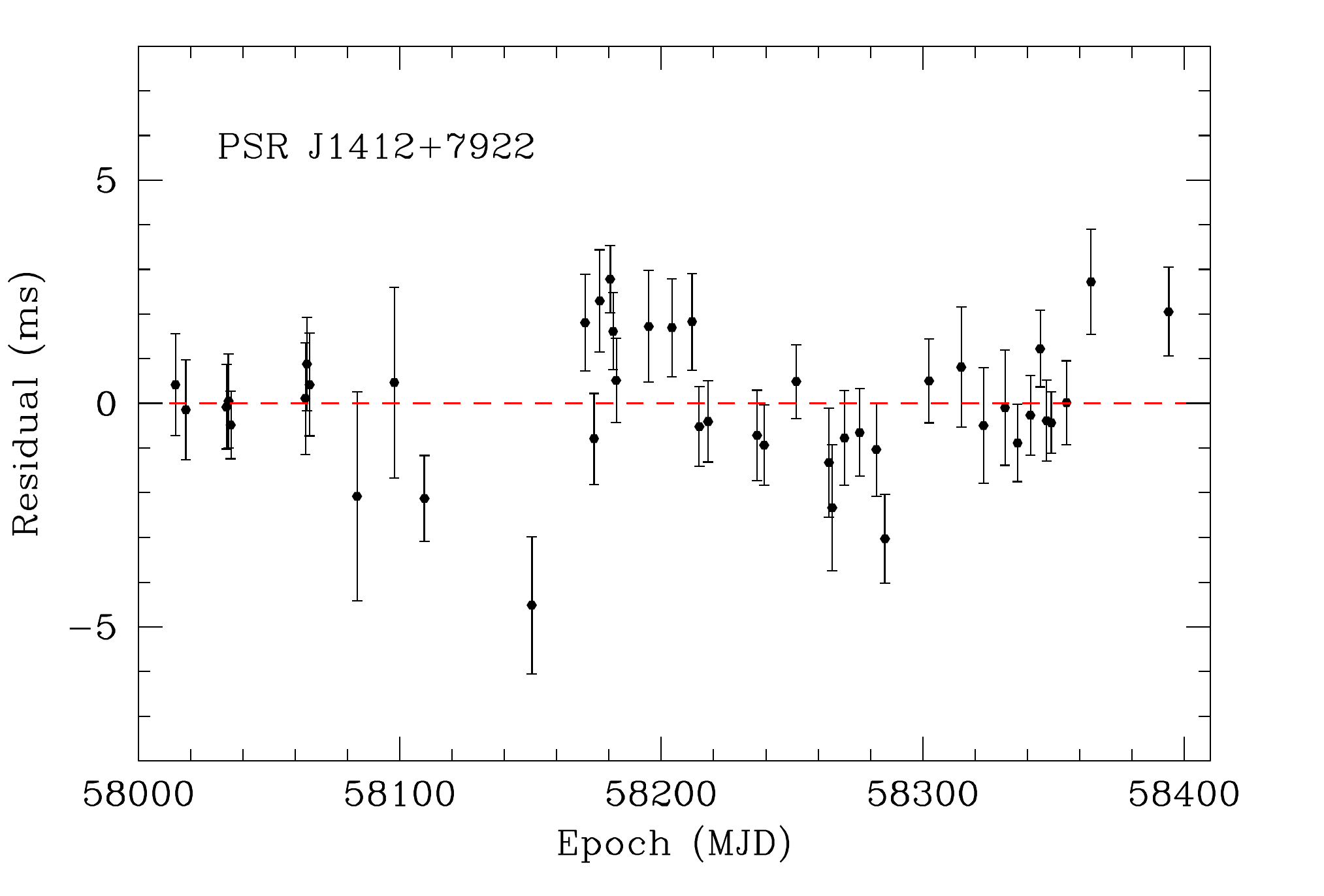}
\includegraphics[clip, trim=1.0cm 11cm 0.0cm 3.0cm, width=0.48\textwidth]{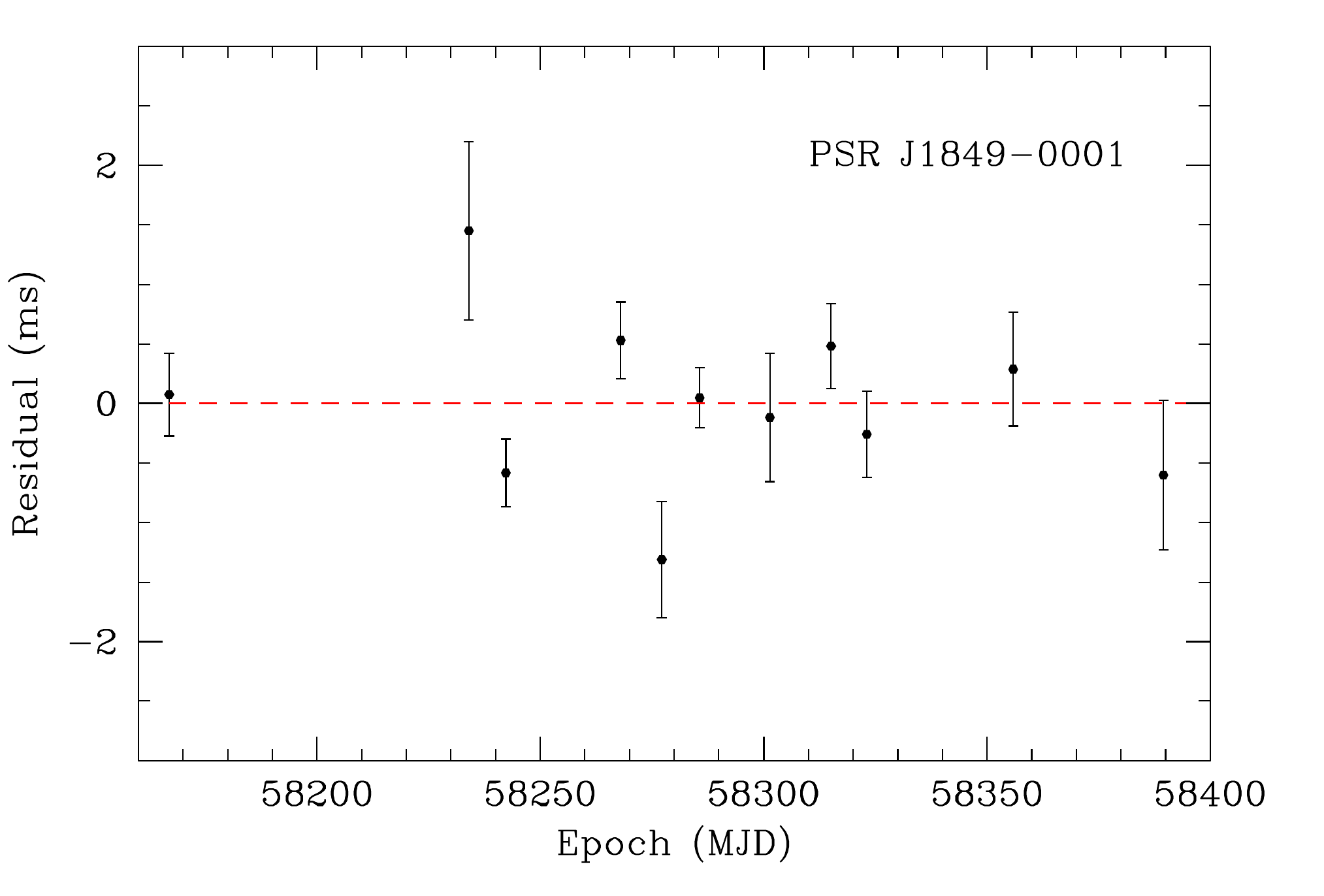}
\caption{\textit{NICER} timing residuals in milliseconds of PSRs J1412+7922 (top) and J1849$-$0001 (bottom) over the full observing time span for each source.}
\label{fig:calvera_toas}
\end{figure}

\begin{deluxetable*}{ll}
\tabletypesize{\small}
\tablecaption{Timing Parameters of PSR J1412$+$7922 (Calvera).}
\tablehead{
\colhead{Parameter}   &
\colhead{Value}
}
\startdata                         
\multicolumn{2}{c}{Assumed parameters\tablenotemark{a}} \\
\hline
R.A. (J2000.0) \dotfill    & $14^{\rm h}12^{\rm m}55.\!^{\rm s}867$ \\
Decl. (J2000.0) \dotfill   & $+79^{\circ}22^{\prime}03.\!^{\prime\prime}895$ \\
$\mu_{\alpha}\cos\delta$ (mas yr$^{-1}$) \dotfill & $-40(30)$ \\
$\mu_{\delta}$ (mas yr$^{-1}$) \dotfill & $-56(21)$ \\
Epoch of astrometric parameters (MJD) \dotfill & $55,449.5$ \\
\hline
\multicolumn{2}{c}{Derived parameters} \\
\hline
Epoch (MJD TDB) \dotfill                  &  $58150.46948$ \\
Period\tablenotemark{b}, $P$ (ms) \dotfill                 & $59.199071070(18)$   \\
Period derivative\tablenotemark{b}, $\dot P$ (s~s$^{-1}$) \dotfill & $3.29134(30) \times 10^{-15}$ \\
Range of dates (MJD) \dotfill                 &  $58014.2-58394.1$  \\
Spin-down luminosity, $\dot E$ (erg s$^{-1}$)\dotfill       &  $6.3\times10^{35}$         \\
Characteristic age, $\tau_c$ (kyr) \dotfill         &  $285$           \\
Surface dipole magnetic field, $B_s$ (G) \dotfill & $4.5\times10^{11}$
\enddata
\tablenotetext{a}{\footnotesize \textit{Chandra} position and proper motion from \cite{2015ApJ...812...61H}.}
\tablenotetext{b}{\footnotesize Tempo2 $1\sigma$ uncertainties given in parentheses.}
\label{tab:j1412_ephem}
\end{deluxetable*}

\begin{deluxetable}{ll}
\tabletypesize{\small}
\tablecaption{Timing Parameters of PSR J1849$-$0001}
\tablehead{
\colhead{Parameter}   &
\colhead{Value}
}
\startdata        
\multicolumn{2}{c}{Assumed parameters\tablenotemark{a}} \\
\hline
R.A. (J2000.0) \dotfill    & $18^{\rm h}49^{\rm m}01.\!^{\rm s}632$ \\
Decl. (J2000.0) \dotfill   & $-00^{\circ}01^{\prime}17.\!^{\prime\prime}45$ \\
\hline
\multicolumn{2}{c}{Derived parameters} \\
\hline
Epoch (MJD TDB) \dotfill                    &  $58239.916286135$ \\
Period\tablenotemark{b}, $P$ (ms) \dotfill                 & $38.5222586295(19)$   \\
Period derivative\tablenotemark{b}, $\dot P$  (s~s$^{-1}$) \dotfill & $1.415734(49) \times 10^{-14}$ \\
Range of dates (MJD) \dotfill                 &  $58166.4-58355.9$  \\
Spin-down luminosity, $\dot E$ (erg s$^{-1}$) \dotfill       & $9.8\times10^{36}$          \\
Characteristic age, $\tau_c$ (kyr) \dotfill         &     $43.1$       \\
Surface dipole magnetic field, $B_s$ (G) \dotfill &  $7.5\times10^{11}$ 
\enddata
\tablenotetext{a}{\footnotesize \textit{Chandra} position from \cite{2015MNRAS.449.3827K}.}
\tablenotetext{b}{\footnotesize Tempo2 $1\sigma$ uncertainties given in parentheses.}
\label{tab:j1849_ephem}
\end{deluxetable}

\section{Timing Analysis} \label{sec:timing}

Data processing and filtering were accomplished using HEASoft 6.24\footnote{\url{https://heasarc.nasa.gov/lheasoft/}} and NICERDAS 2018-04-13\_V004. The event data were first filtered to exclude any portions of the exposure accumulated during passages of the ISS through the South Atlantic Anomaly (SAA). For the purposes of producing event lists optimized for timing analysis, the data were further screened for instances of elevated count rates on a per detector basis, which revealed that detector 34 frequently showed count rates well above the median count rate of all active detectors. After excluding all events from this ``hot'' detector, a final pass was made to excise any time intervals of enhanced background affecting all detectors. This was done by constructing a light curve binned at 8 seconds and removing the time bins in which the count rate was in excess of 4.5 and 5.0 counts s$^{-1}$ for PSR J1412$+$7922 and PSR J1849$-$0001, respectively, in the full \textit{NICER} band (0.25--12 keV). These relatively high thresholds were selected in order to remove only periods strongly contaminated by background flaring.  Using these filtering criteria, we obtained total clean exposures of $371.8$ ks and $60.4$ ks for the two sources for use in pulse timing analyses.

Since PSR J1412$+$7922 is known to be a soft thermal source, only events in the 0.3--3 keV range were considered in the timing analysis presented below. PSR J1849$-$0001, on the other hand, suffers from strong interstellar absorption ($N_H\approx 10^{22}$ cm$^{-2}$) so nearly all source emission is above $\sim$1 keV. Based on this, we selected only events in the 1--6 keV range for timing purposes.  To correct for the telescope motion and make the transformation between Terrestrial Time (TT), used for the \textit{NICER} event time stamps, and Barycentric Dynamical Time (TDB) we assumed the JPL DE421 solar system ephemeris.   For both sources, as free pulsar parameters in the timing model we only consider the period $P$ and period derivative $\dot{P}$. The sky positions for both pulsars were fixed at the values previously measured to sub-arcsecond precision with \textit{Chandra}. Although PSR J1412$+$7922 has a measured proper motion, we only use it to correct for the resulting shift in position over time and do not fit for it.
 
Following standard pulsar timing procedures, we started by combining exposures taken on adjacent days to produce a single pulse time of arrival (TOA) measurement. The data were combined such that there is sufficient exposure to confidently detect the pulsar (typically 6-8 ks for both pulsars), while restricting the time span of each TOA to less then seven days. The ObsID groupings of these TOAs are listed in the last column of Tables \ref{tab:j1412_obs}1 and \ref{tab:j1849_obs}2. This resulted in 44 and 11 \textit{NICER} TOAs for PSR J1412$+$7922 and PSR J1849$-$0001, respectively. For both pulsars, we searched each ObsID grouping for pulsations around the known spin parameters to produce a folded and binned pulse profile at the detected periodicity. We fitted the profiles from the TOA where the pulsar is detected with the highest statistical significance (TOAs \#16 and \#3 for the two pulsars, respectively) as determined by the H-test \citep{1989A&A...221..180D}, with a single-peaked symmetric Gaussian template to determine the fiducial phase corresponding to the peak of the pulse (which we designate as $\phi=0$). This template profile was used to generate a set of TOAs that were fit with a timing model using Tempo2 \citep{2006MNRAS.369..655H}.  In a final iteration, an improved template was generated using the entire refolded data set and with energy cuts that maximize the pulsation detection significance (0.37--1.97 keV for PSR J1412$+$7922 and 1.89--6 keV for PSR J1849$-$0001), photon phases were reassigned using the improved solution, and new TOAs were produced and refit to arrive at the final timing solution summarized in Tables \ref{tab:j1412_ephem} and \ref{tab:j1849_ephem}. The root mean square post-fit timing residuals for the two pulsars are 1.36 ms and 0.525 ms, respectively, which are the expected levels given the broad pulses. Adding a second period derivative ($\ddot{P}$) in the timing model for either pulsar does not result in a statistically significant improvement in the residuals and its value is consistent with zero.

\begin{figure}
\centering
\includegraphics[clip, trim=1.5cm 12.5cm 0.0cm 3.0cm, width=0.45\textwidth]{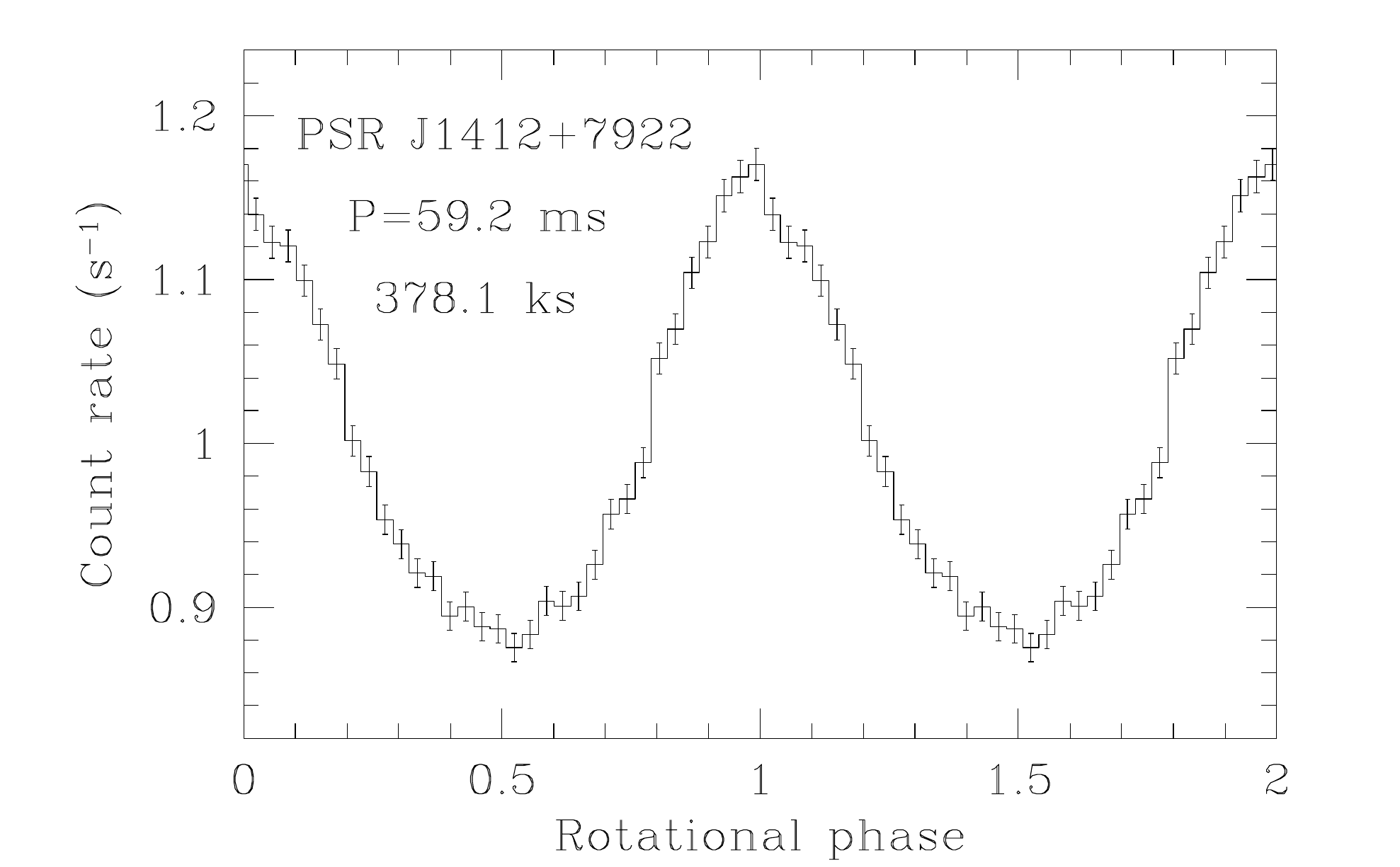}
\includegraphics[clip, trim=1.5cm 12.5cm 0.0cm 3.0cm, width=0.45\textwidth]{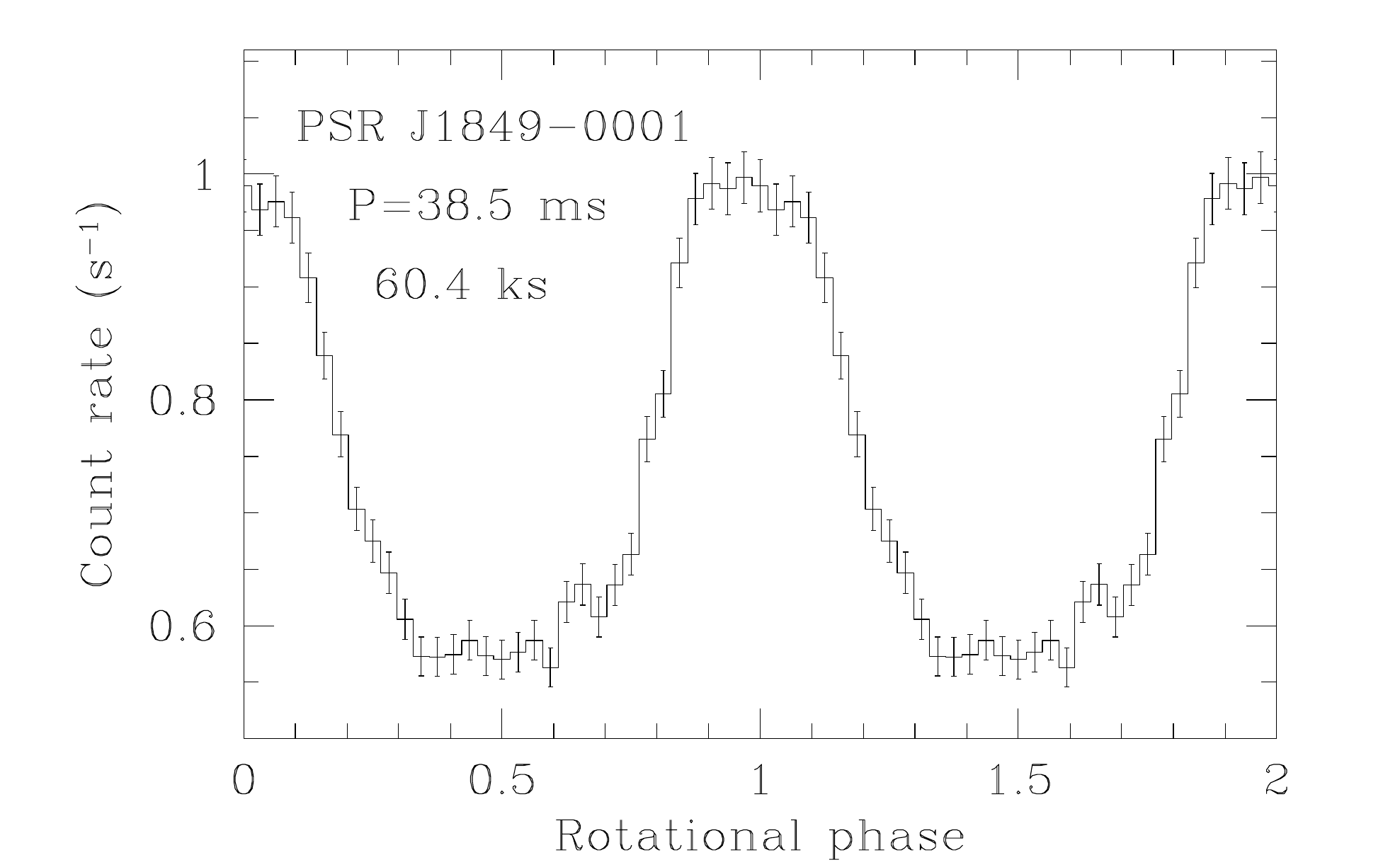}
\caption{Pulse profiles of PSRs J1412+7922 in the 0.37--1.97 keV band (top) and J1849$-$0001 over 1.89--6 keV (bottom), folded at the ephemerides reported in Tables \ref{tab:j1412_ephem} and \ref{tab:j1849_ephem} using all available \textit{NICER} data.}
\label{fig:calvera_profile}
\end{figure}

The \textit{NICER} timing residuals for PSR J1412$+$7922 and PSR J1849$-$0001 are shown in Figure~\ref{fig:calvera_toas}, while the {NICER} pulse profiles of both pulsars folded using the new timing solutions are presented in Figure~\ref{fig:calvera_profile}. The derived spin parameters ($P$ and $\dot{P}$) for both pulsars are fully consistent with previous measurements but with substantially reduced uncertainties, with the values of $\dot{P}$ in particular being improved by two orders of magnitude. We note that some of the scatter evident in the TOA residuals may be due to unpredictable ``timing noise'' commonly seen in young pulsars.

\begin{figure}
\includegraphics[width=0.49\textwidth]{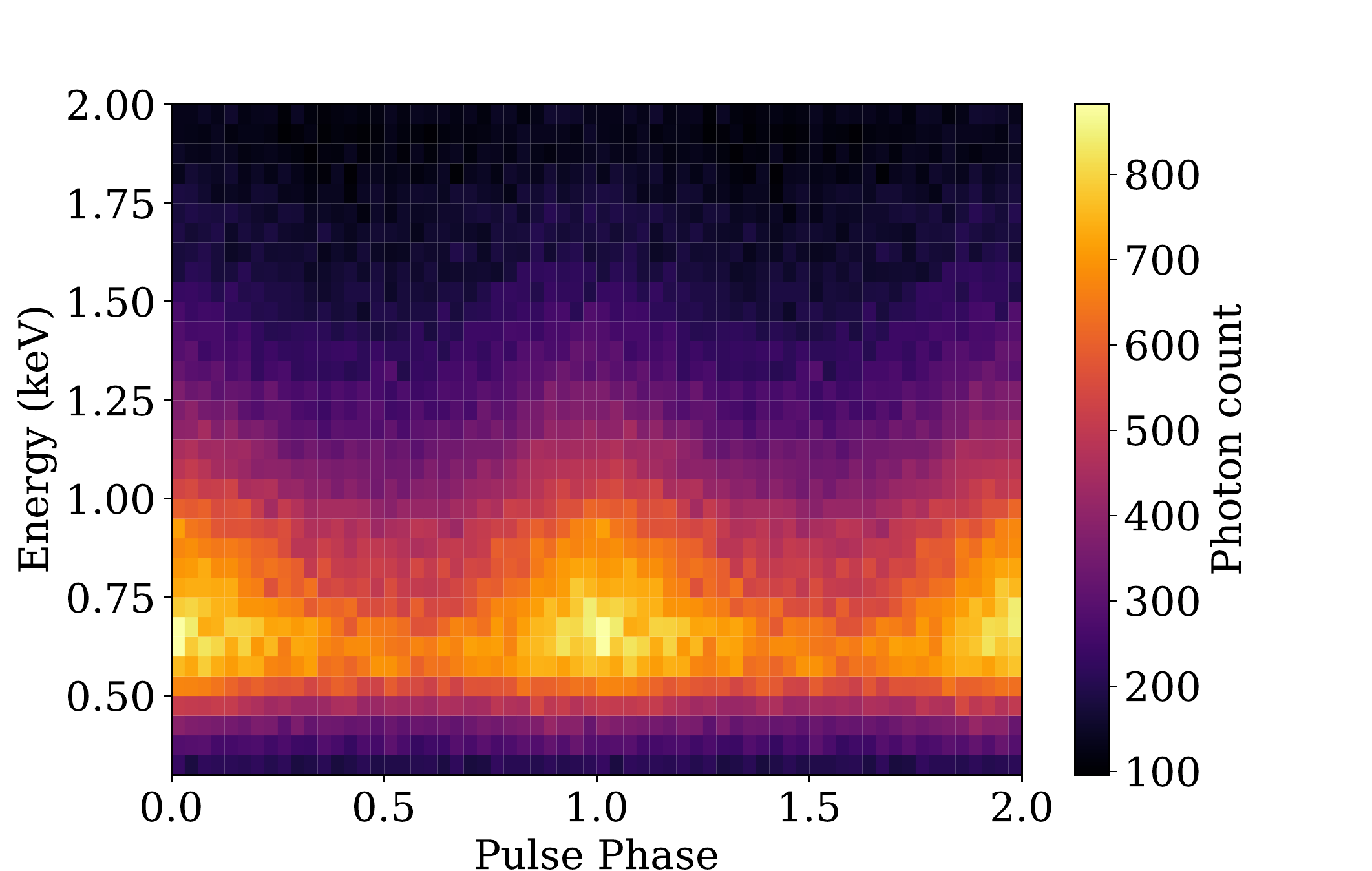}
\includegraphics[width=0.49\textwidth]{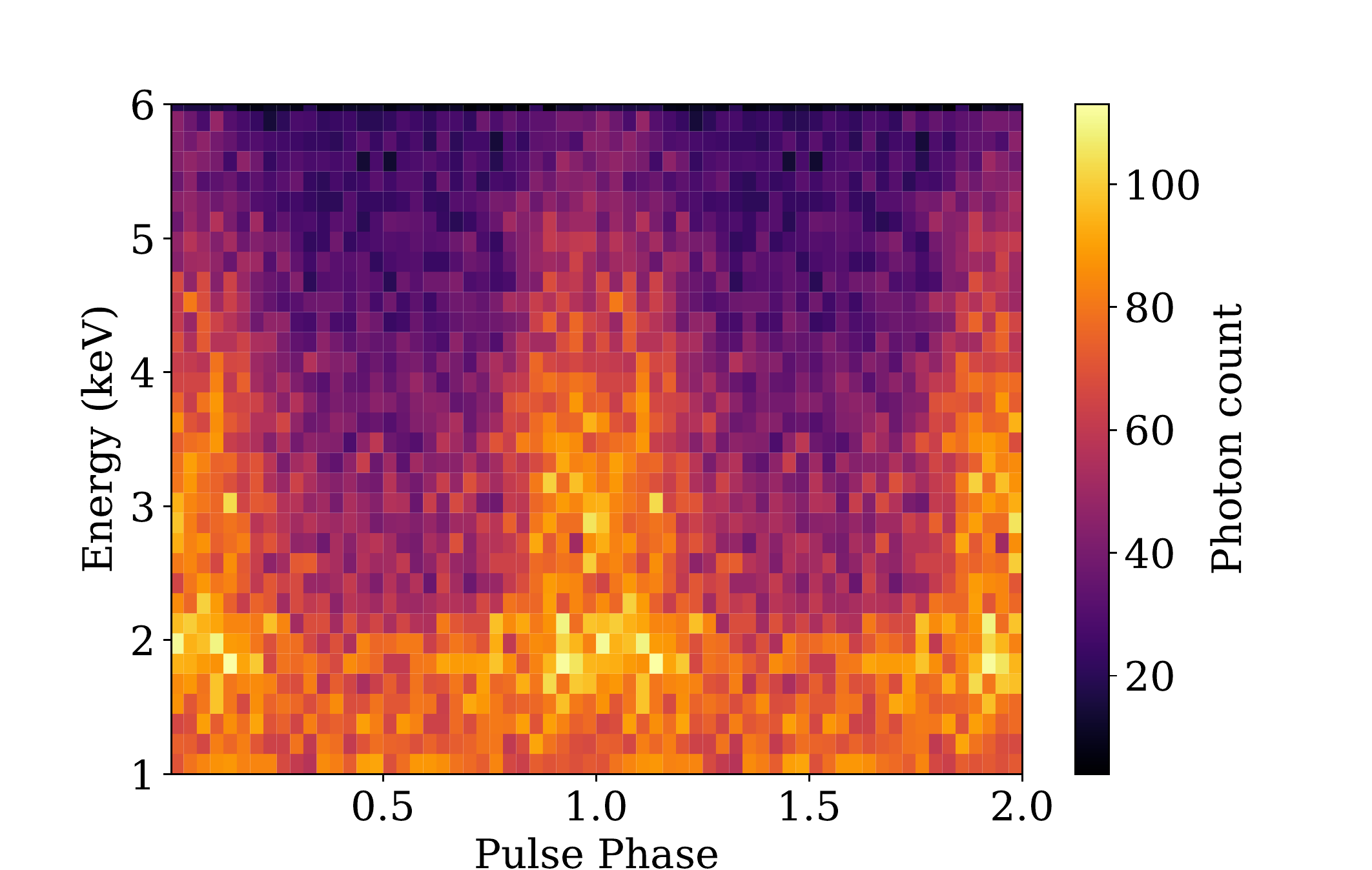}
\caption{Photon counts as a function of pulse phase and energy for PSRs J1412+7922 (top) and J1849$-$0001 (bottom) folded at the ephemerides reported in Tables 1 and 2 using all available \textit{NICER} data.}
\label{fig:calvera_2d}
\end{figure}


\section{Spectral analysis} \label{sec:spectra}

Figure~\ref{fig:calvera_2d} shows two-dimensional count maps versus pulse phase and energy for the two pulsars. These plots illustrate that PSR~J1412+7922 is bright only in soft X-rays, while PSR~J1849$-$0001 is relatively hard and detected in the higher energy band up to at least $\sim$6~keV. We carried out \textit{NICER} spectral analyses 
with the calibration database (CALDB) version 20181105 and gain solution version {\tt optmv7}.
Cleaned events were extracted from Good Time Intervals (GTIs) defined by the standard filtering criteria, together with additional constraints based on a space-weather background model developed within the \textit{NICER} team:  $\textrm{KP}<5$, where $\textrm{KP}$ corresponds to the $K_p$ geomagnetic activity index; $\textrm{COR\_SAX} > 1.914\times \textrm{KP}^{0.684}+0.25$, where COR\_SAX is the magnetic cut-off rigidity (in units of GeV/c); and $\textrm{FPM\_UNDERONLY\_COUNT}<200$, where $\textrm{FPM\_UNDERONLY\_COUNT}$ represents the rate of ``undershoot'' resets per enabled FPM per second, a measure of optical light-loading. 
In the following spectral studies, we do not use the filtering criteria to exclude time bins with high count rates described in \S2. The corresponding background spectra are estimated using the space-weather background model and observations of seven ``blank sky'' fields adopted from \textit{RXTE} studies \citep{2006ApJS..163..401J}.
Three FPMs (DET\_IDs 14, 34, and 54) are excluded from our spectral analyses because they sometimes exhibit higher background rates in the low-energy band compared to the other FPMs. 
We used a standard response matrix file (RMF; version 1.01) in the CALDB and an ancillary response file (ARF) scaled to account for the three excluded detector modules.

\begin{figure}
\includegraphics[clip,width=0.46\textwidth]{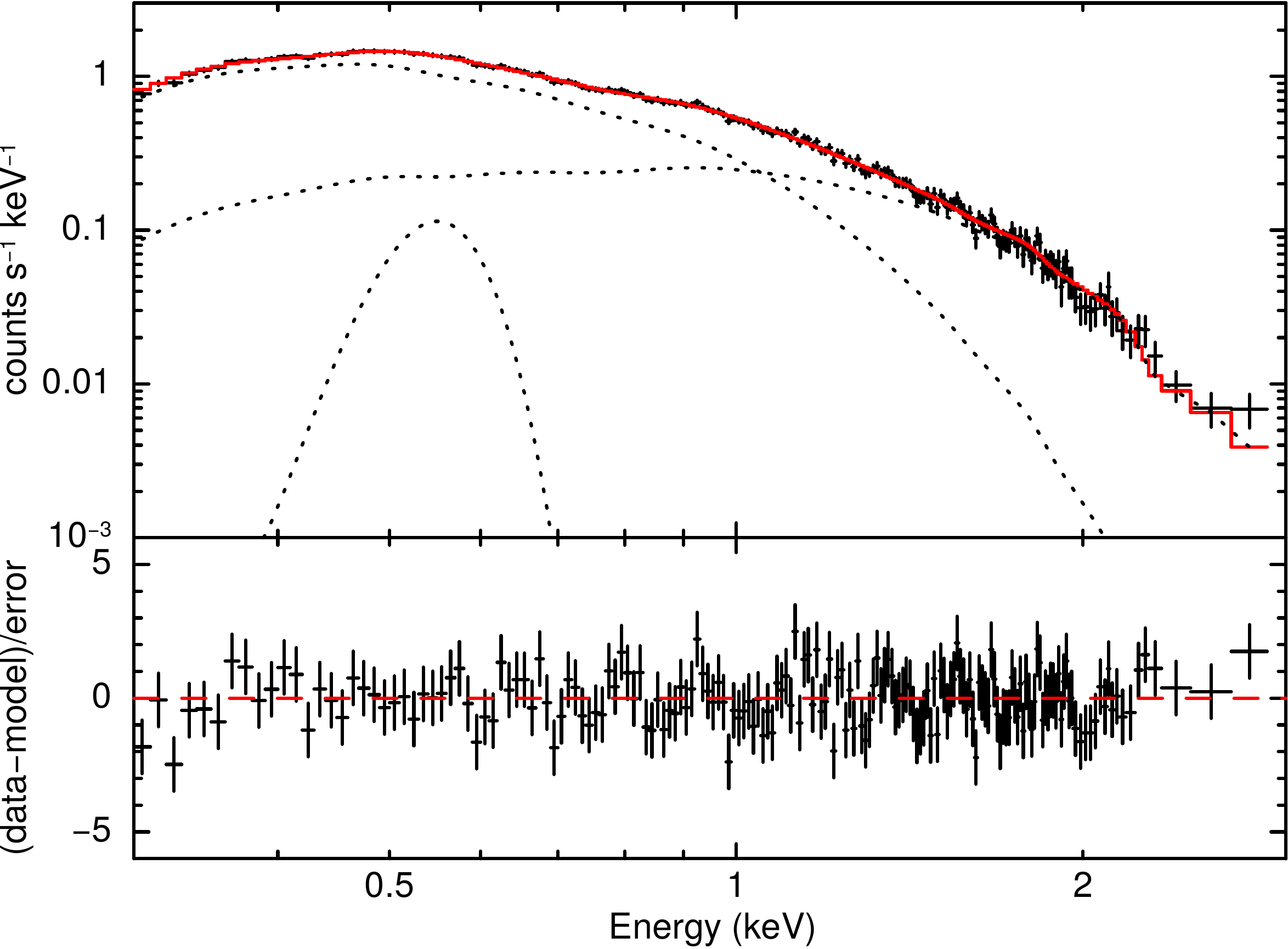}
\includegraphics[clip,width=0.46\textwidth]{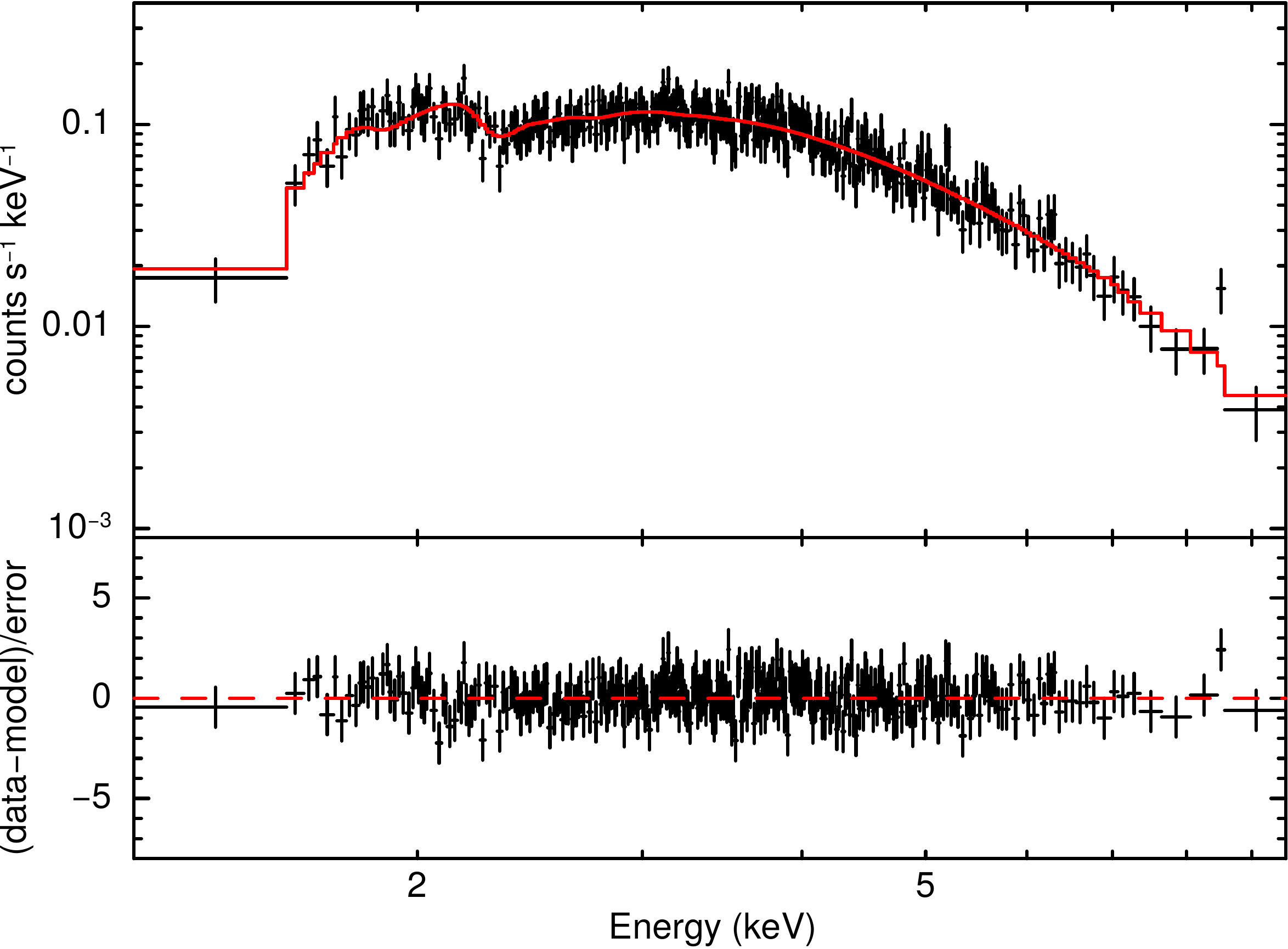}
\caption{Spectral fitting of PSR~J1412$+$7922 (Calvera, top panel) and PSR~J1849$-$0001 (bottom panel). The two-temperature blackbody model is used for  PSR~J1412$+$7922, while the single power-law model is shown for PSR~J1849$-$0001. See Tables~\ref{tab:j1412_spectra} and \ref{tab:j1849_spectra} for the best fit parameters.}
\label{fig:spectra}
\end{figure}

We measure background-subtracted source count rates of 0.94 and 0.41 counts~s$^{-1}$ for PSR~J1412$+$7922 (0.3--3.0~keV) and PSR~J1849$-$0001 (1--10~keV), respectively. Derived spectra are binned so that individual bins have either $>4 \sigma$ detection significance or up to 50 counts and 100 counts for PSR~J1412$+$7922 and  PSR~J1849$-$0001, respectively.
Figure~\ref{fig:spectra} shows the data and best-fit spectral models for these two pulsars. 

\begin{deluxetable*}{llcc}
\tabletypesize{\small}
\tablecaption{Spectral Fit Parameters of PSR~J1412$+$7922 (Calvera)}
\tablewidth{0pc}
\tablehead{
\colhead{Component} &
\colhead{Parameter} &
\colhead{BB+PL} &
\colhead{BB+BB}
}
\startdata
\texttt{tbabs}          & $N_{\rm H}$ ($10^{20}\mbox{ cm$^{-2}$}$) & $5.14\pm 0.31$ & $0.29\pm 0.14$ \\
\texttt{pegpwrlw} (PL)  & $\Gamma$ & $2.99 \pm 0.06$ & \nodata \\
                        & Norm. ($10^{-12}$~ergs~s$^{-1}$~cm$^{-2}$) \tablenotemark{a}  &  0.11$\pm$0.01   & \nodata \\
\texttt{bbodyrad} (BB) & $kT_1$ (keV) & $0.205 \pm 0.003$ & $0.154 \pm 0.004$ \\
                        & $R_1$ (km) \tablenotemark{b} & $1.01^{+0.04}_{-0.03}$ & $2.21_{-0.07}^{+0.08}$ \\ 
\texttt{bbodyrad} (BB) & $kT_2$ (keV) & \nodata & $0.319_{-0.012}^{+0.013}$ \\
                        & $R_2$ (km) \tablenotemark{b} & \nodata & $0.37\pm 0.04$ \\
\texttt{gabs}           & $E_{\rm c}$ (keV) & $0.78\pm 0.02$ & $0.76\pm 0.01$ \\
                        & $\sigma$ (keV) & $0.05 \pm 0.02$  & $0.08\pm 0.01$ \\
                        & strength\tablenotemark{c}  & $0.008\pm 0.003$ & $0.03\pm 0.01$ \\
\hline
\multicolumn{2}{c}{$F_{0.3-3\,{\rm keV}}^{\rm abs}$ ($10^{-13}\mbox{ erg cm$^{-2}$ s$^{-1}$}$)} & $9.5\pm 0.1$ & $9.4\pm 0.2$ \\
\multicolumn{2}{c}{$\chi^2$/dof} & 165.5/173 & 183.8/177
\enddata
\tablenotetext{a}{\footnotesize The normalization of the \texttt{pegpwrlw} model is given in units of 2--10 keV unabsorbed flux.}
\tablenotetext{b}{\footnotesize Distance of 2 kpc assumed.}
\tablenotetext{c}{\footnotesize The line strength is defined by $\sigma\tau\sqrt{2\pi}$, where $\tau$ is the optical depth.}
\label{tab:j1412_spectra}
\end{deluxetable*}

The soft X-ray bright source PSR~J1412$+$7922 was detected below $\sim$2~keV (top panel of Figure~\ref{fig:spectra}). We performed fits of its spectrum in the 0.22--2.1~keV band with models that include photoelectric absorption (\texttt{tbabs} in XSPEC; \citealt{2000ApJ...542..914W}) and two additional components, either two blackbodies (\texttt{bbodyrad}$+$\texttt{bbodyrad}) or a blackbody plus power-law (\texttt{bbodyrad}$+$\texttt{pegpwrlw}) because a single blackbody model can not reproduce the data; we note that \citet{2016ApJ...831..112S} obtain a good fit to rotation phase-averaged \textit{Chandra} and \textit{XMM-Newton} spectra using a single atmosphere model. Both sets of spectral models show fit residuals that can be modeled by an emission feature at $\sim$0.55~keV and an absorption feature at $\sim$0.76~keV. The former is thought to be a foreground feature due to solar wind charge exchange or a local hot bubble along the line of sight, but it could be related to a 0.5 keV emission feature reported in the \textit{Chandra} spectrum \citep{2009ApJ...705..391S}. We added a Gaussian emission model (\texttt{gaussian}) to take it into account.
The latter absorption feature was previously seen in an \textit{XMM-Newton} spectrum \citep{2016ApJ...831..112S}, and we find it in our \textit{NICER} data to be phase-dependent with an energy shift over the pulse rotation. To account for it, we added a Gaussian absorption model (\texttt{gabs} in XSPEC). The best fit parameters are summarized in Table~\ref{tab:j1412_spectra}.
Detailed phase-resolved spectral analysis with more realistic spectral models is the subject of ongoing work.

The $\sim$1.5--9~keV spectrum of PSR~J1849$-$0001 (bottom panel of Figure~\ref{fig:spectra}) is well-fit by a photoelectric absorption model (\texttt{tbabs}) and either a single blackbody or power-law model. The best-fit parameters are listed in Table~\ref{tab:j1849_spectra}. We favor the power-law model, which was used in analyzing previous \textit{RXTE} and \textit{XMM-Newton} observations \citep{2011ApJ...729L..16G,2015MNRAS.449.3827K,2018arXiv180204833V}, because it is a better fit than the blackbody model, with an improvement of $\triangle \chi^2=-31.9$,
and the very high blackbody temperature is likely unrealistic.
Our derived X-ray flux ($F_{2-10}^{\rm unabs}=(6.8\pm 0.1)\times10^{-12}\mbox{ erg cm$^{-2}$ s$^{-1}$}$), 
absorption ($N_{\rm H}=6.2\times 10^{22}\mbox{ cm$^{-2}$}$), and photon index ($\Gamma=1.54$) are different from previous \textit{XMM-Newton} values ($F_{2-10}^{\rm unabs}\approx 4.8\times10^{-12}\mbox{ erg cm$^{-2}$ s$^{-1}$}$, $N_{\rm H}=4.5\times 10^{22}\mbox{ cm$^{-2}$}$, and $\Gamma=1.3$; \citealt{2015MNRAS.449.3827K}).  However, as discussed in Section~\ref{sec:intro}, there is notable diffuse emission around the pulsar due to its PWN, and the non-imaging detectors of \textit{NICER} cannot resolve these two components.

\begin{deluxetable*}{llcc}
\tabletypesize{\small}
\tablecaption{Spectral Fit Parameters of PSR J1849$-$0001}
\tablehead{
\colhead{Component} &
\colhead{Parameter} &
\colhead{PL} &
\colhead{BB}
}
\startdata
\texttt{tbabs}          & $N_{\rm H}$ ($10^{22}\mbox{ cm$^{-2}$}$) & $6.23_{-0.26}^{+0.27}$ & $3.23_{-0.16}^{+0.17}$ \\
\texttt{pegpwrlw} (PL)  & $\Gamma$ & $1.54\pm0.07$ & \nodata \\
                        & Norm. ($10^{-12}$~ergs~s$^{-1}$~cm$^{-2}$)  \tablenotemark{a}  &  $6.75^{+0.12}_{-0.11}$   & \nodata \\
\texttt{bbodyrad} (BB) & $kT$ (keV) & \nodata & $1.43\pm0.04$ \\
                        & $R$(km) \tablenotemark{b}  & \nodata & $0.254\pm0.011$ \\ 
\hline
\multicolumn{2}{c}{$F_{0.3-3\,{\rm keV}}^{\rm abs}$ ($10^{-12}\mbox{ erg cm$^{-2}$ s$^{-1}$}$)} &  $5.14_{-0.15}^{+0.11}$ & $4.30_{-0.10}^{+0.09}$ \\
\multicolumn{2}{c}{$\chi^2$/dof} & 225.4/309 & 257.3/309
\enddata
\tablenotetext{\tiny a}{\footnotesize The normalization of the \texttt{pegpwrlw} model is given in units of 2--10 keV unabsorbed flux.}
\tablenotetext{\tiny b}{\footnotesize Distance of 7 kpc assumed.}
\label{tab:j1849_spectra}
\end{deluxetable*}

\section{Conclusions}\label{sec:conclusions}
We presented new X-ray timing and spectroscopic analyses of the X-ray-only pulsars PSRs J1412$+$7922 and J1849$-$0001 based on \textit{NICER} data. We obtain phase-connected timing solutions spanning 380 and 223 days for the two pulsars, respectively. The \textit{NICER} timing campaigns enable a two orders of magnitude improvement in the spin-down rate measurements compared to previous studies of the two pulsars.

The LIGO Scientific Collaboration and Virgo Collaboration perform targeted searches of known pulsars whose accurately measured position and spin properties reduce the parameter space for GW searches.  With an accurate timing model, phase-coherent GW searches can be performed over long periods of time, thereby maximizing the potential for detection.  Past targeted searches primarily used timing models obtained by radio telescopes and \textit{Fermi} to place constraints on GW emission from some 200 radio and $\gamma$-ray pulsars (\citealt{2017ApJ...839...12A}; see also \citealt{2014ApJ...785..119A,2017PhRvD..96l2006A}).

The GW strain amplitude $h_0(\nu_{\rm GW})$ that is produced at GW frequency $\nu_{\rm GW}$ ($=2\nu$, where $\nu=1/P$) by a quadrupolar mass deformation with ellipticity $\varepsilon\equiv\left|I_{xx}-I_{yy}\right|/I_{zz}$, where $I_{xx}$, $I_{yy}$, and $I_{zz}$ are the triaxial components of the stellar moment of inertia, is (see, e.g., \citealt{2014ApJ...785..119A})
\begin{eqnarray}
h_0 &=& \frac{16\pi^2G}{c^4}\frac{\varepsilon I_{zz}}{D}\nu^2 \nonumber\\
&=& 4.23\times 10^{-26}\left(\frac{1\mbox{ kpc}}{D}\right)\left(\frac{\varepsilon}{10^{-4}}\right)\left(\frac{\nu}{10\mbox{ Hz}}\right)^2. \label{eq:h0}
\end{eqnarray}
For the dominant r-mode fluid oscillation of dimensionless amplitude $\alpha$, $\nu_{\rm GW}\approx 4\nu/3$ \citep{2014MNRAS.442.1786A,2015PhRvD..91b4001I,2017PhRvD..95f4060J} and the equivalent GW strain amplitude $h_0$ (see, e.g., \citealt{2010PhRvD..82j4002O}) is
\begin{equation}
h_0\sim 5.36\times 10^{-26}\alpha\left(\frac{1\mbox{ kpc}}{D}\right)\left(\frac{\nu}{10\mbox{ Hz}}\right)^3. \label{eq:h0r}
\end{equation}

An upper bound on the GW strain that can be produced by a pulsar with known $\nu$ and $\dot{\nu}$ ($=-\dot{P}/P^2$) is obtained by assuming that the pulsar rotational energy loss is due entirely to GW emission (i.e., neglecting electromagnetic losses).  In such an idealized case, the ``spin-down limit'' on GW strain for a quadrupolar mass deformation is
\begin{eqnarray}
h_{\rm sd} &=& \left(-\frac{5G}{2c^3}\frac{I_{zz}}{d^2}\frac{\dot{\nu}}{\nu}\right)^{1/2} \nonumber\\
&=& 2.55\times 10^{-25}\left(\frac{1\mbox{ kpc}}{D}\right)  \nonumber\\
&&\times \left(\frac{10\mbox{ Hz}}{\nu}\right)^{1/2}\left(\frac{-\dot{\nu}}{10^{-12}\mbox{ Hz s$^{-1}$}}\right)^{1/2}, \label{eq:hsd}
\end{eqnarray}
while the spin-down limit for a r-mode fluid oscillation is 3/2 that given by equation~(\ref{eq:hsd}) \citep{2010PhRvD..82j4002O}.  Thus, as GW searches become more sensitive (pushing the measured $h_0$ to lower values, such that $h_0<h_{\rm sd}$), the constraint on  ellipticity improves as
\begin{equation}
\varepsilon=6.03\times 10^{-4}\!\left(\frac{10\mbox{ Hz}}{\nu}\right)^{5/2}\!\!\!\left(\frac{-\dot{\nu}}{10^{-12}\mbox{ Hz s$^{-1}$}}\right)^{1/2}\!\!\!\left(\frac{h_0}{h_{\rm sd}}\right). \label{eq:ellip}
\end{equation}

Similarly, the constraint on r-mode amplitude is
\begin{equation}
\alpha=7\left(\frac{10\mbox{ Hz}}{\nu}\right)^{7/2} \left(\frac{-\dot{\nu}}{10^{-12}\mbox{ Hz s$^{-1}$}}\right)^{1/2}\left(\frac{h_0}{h_{\rm sd}}\right). \label{eq:alpha}
\end{equation}

For an assumed (maximum) distance of 2~kpc to PSR~J1412+7922 \citep{2013ApJ...778..120H,2015ApJ...812...61H}, we find that the GW spindown limit is $h_{\rm sd}=9.51\times 10^{-26}$.  For an assumed distance of 7~kpc to PSR~J1849$-$0001 \citep{2011ApJ...729L..16G}, we find $h_{\rm sd}=6.98\times 10^{-26}$.  These $h_{\rm sd}$ can be approximately compared to the estimated GW strain sensitivity $h_0(\nu_{\rm GW})$ of the LIGO O1 run given in \citet{2017ApJ...839...12A}; in particular, $h_0(23\mbox{ Hz})\approx 3\times 10^{-25}$, $h_0(34\mbox{ Hz})\approx 9\times 10^{-26}$, and $h_0(52\mbox{ Hz})\approx 4\times 10^{-26}$.  Thus O1 data is at or two times the spindown limit for PSR~J1412+7922 and is lower than the spindown limit for PSR~J1849$-$0001.  The timing models obtained from \textit{NICER} data and presented here enable future GW observations to obtain constraints for each X-ray pulsar.  For example, a GW search at O1 sensitivity would limit the ellipticity of both PSR~J1412+7922 and PSR~J1849$-$0001 at $\varepsilon<1\times 10^{-4}$ and, for PSR~J1849$-$0001, r-mode amplitude at $\alpha<0.7$ and GW contribution to the pulsar's rotational energy loss at $<30\%$.  Of course, future GW searches will have greatly improved sensitivities (see, e.g., \citealt{2018LRR....21....3A}), such that much stronger limits will be achieved.  While neither pulsar has been seen to glitch, glitch activity is known to correlate with age (e.g., \citealt{2017A&A...608A.131F}).  Young pulsars also tend to exhibit spin fluctuations (timing noise),  which only contemporaneous observations can track.  Thus it is important to continue monitoring these relatively young pulsars in X-rays (and other X-ray-only pulsars) with \textit{NICER} to maintain the accuracy of each timing model in order to complement GW searches.

\appendix

\section{Log of \textit{NICER} Observations of PSR J1412$+$7922 and PSR J1849$-$0001}
The set of \textit{NICER} observations of PSR J1412$+$7922 and PSR J1849$-$0001 are summarized in Tables \ref{tab:j1412_obs}1 and \ref{tab:j1849_obs}2, respectively. The tables list the ObsIDs, the start time of the exposure in UTC units, the good exposure time after excluding intervals during SAA passage, and the TOA number in which the observation was included for the timing analysis.

\acknowledgments

S.B. thanks E.V.~Gotthelf for numerous insightful discussions concerning X-ray timing of pulsars. W.C.G.H. thanks G.~Woan for discussions and K. Riles for comments. This work was supported in part by NASA through the \textit{NICER} mission and the Astrophysics Explorers Program. W.C.G.H. acknowledges partial support through grant ST/R00045X/1 from Science and Technology Facilities Council (STFC) in the UK. S.G. acknowledges the support of the Centre National d'Etudes Spatiales (CNES). This research has made use of data and software provided by the High Energy Astrophysics Science Archive Research Center (HEASARC), which is a service of the Astrophysics Science Division at NASA/GSFC and the High Energy Astrophysics Division of the Smithsonian Astrophysical Observatory.  We acknowledge extensive use of the NASA Abstract Database Service (ADS) and the ArXiv.

\facilities{\textit{NICER}}
%


\software{Tempo2 \citep{2006MNRAS.369..655H}}


\bibliography{slavko_master_refs.bib}


\startlongtable
\begin{deluxetable}{clrc}
\renewcommand\thetable{A1}
\tabletypesize{\footnotesize}
\label{tab:j1412_obs}
\tablecaption{Log of \textit{NICER} observations of PSR J1412$+$7922}
\tablecolumns{4}
\tablewidth{0pc}
\tablehead{
\colhead{Observation}   &
\colhead{Start time} &
\colhead{Exposure} & 
\colhead{TOA}\\
\colhead{ID}   &
\colhead{(UTC)} &
\colhead{(s)} & 
\colhead{Number}
}
\startdata
1020290101	&	2017-09-15T00:03:43	&	2329	&	1	\\
1020290102	&	2017-09-18T05:03:20	&	4072	&	1	\\
1020290103	&	2017-09-19T07:16:00	&	726	&	1	\\
1020290104	&	2017-09-20T03:19:20	&	1256	&	1	\\
1020290105	&	2017-09-21T08:56:55	&	2521	&	2	\\
1020290106	&	2017-09-22T04:42:58	&	5693	&	2	\\
1020290107	&	2017-10-07T07:29:20	&	10070	&	3	\\
1020290108	&	2017-10-08T02:02:40	&	10399	&	4	\\
1020290109	&	2017-10-08T23:34:40	&	18025	&	5	\\
1020290110	&	2017-10-10T00:16:17	&	6426	&	6	\\
1020290111	&	2017-11-06T22:27:00	&	221	&	6	\\
1020290112	&	2017-11-06T23:51:04	&	7983	&	7	\\
1020290113	&	2017-11-08T00:42:00	&	6540	&	8	\\
1020290114	&	2017-11-25T17:05:40	&	268	&	9	\\
1020290115	&	2017-11-26T16:06:40	&	1006	&	9	\\
1020290116	&	2017-11-26T23:50:00	&	533	&	9	\\
1020290117	&	2017-12-10T08:26:04	&	962	&	10	\\
1020290118	&	2017-12-11T04:33:00	&	436	&	10	\\
1020290119	&	2017-12-21T19:27:36	&	338	&	11	\\
1020290120	&	2017-12-22T02:54:17	&	9258	&	11	\\
1020290121	&	2018-02-01T11:05:14	&	3650	&	12	\\
1020290122	&	2018-02-21T00:23:20	&	6154	&	13	\\
1020290123	&	2018-02-22T07:17:33	&	2214	&	13	\\
1020290124	&	2018-02-23T09:28:40	&	3154	&	14	\\
1020290125	&	2018-02-24T16:25:00	&	2147	&	14	\\
1020290126	&	2018-02-25T03:15:00	&	1040	&	14	\\
1020290127	&	2018-02-26T00:50:00	&	2817	&	14	\\
1020290128	&	2018-02-27T05:50:40	&	7033	&	15	\\
1020290129	&	2018-03-03T01:11:40	&	14462	&	16	\\
1020290130	&	2018-03-04T03:10:00	&	11346	&	17	\\
1020290131	&	2018-03-05T01:06:00	&	7839	&	18	\\
1020290132	&	2018-03-06T00:15:40	&	1796	&	18	\\
1020290133	&	2018-03-17T22:57:33	&	821	&	19	\\
1020290134	&	2018-03-18T00:30:10	&	5450	&	19	\\
1020290135	&	2018-03-26T01:34:40	&	2163	&	20	\\
1020290136	&	2018-03-27T00:42:16	&	4741	&	20	\\
1020290137	&	2018-03-31T23:41:00	&	461	&	21	\\
1020290138	&	2018-04-01T18:14:40	&	480	&	21	\\
1020290139	&	2018-04-02T00:24:00	&	3198	&	21	\\
1020290140	&	2018-04-03T18:06:00	&	629	&	21	\\
1020290141	&	2018-04-04T17:05:00	&	778	&	21	\\
1020290142	&	2018-04-05T00:58:20	&	2915	&	21	\\
1020290143	&	2018-04-06T00:06:40	&	8892	&	22	\\
1020290144	&	2018-04-07T00:49:40	&	6788	&	23	\\
1020290145	&	2018-04-08T02:49:20	&	1514	&	23	\\
1020290146	&	2018-04-09T11:10:20	&	818	&	23	\\
1020290147	&	2018-04-10T19:37:38	&	856	&	23	\\
1020290148	&	2018-04-11T11:07:37	&	419	&	23	\\
1020290149	&	2018-04-12T07:17:20	&	374	&	23	\\
1020290150	&	2018-04-27T00:34:39	&	6866	&	24	\\
1020290151	&	2018-04-28T13:31:40	&	329	&	24	\\
1020290152	&	2018-05-01T00:17:37	&	12147	&	25	\\
1020290153	&	2018-05-13T05:23:17	&	14394	&	26	\\
1020290154	&	2018-05-22T22:17:37	&	478	&	27	\\
1020290155	&	2018-05-23T01:22:58	&	1342	&	27	\\
1020290156	&	2018-05-24T02:04:54	&	1120	&	27	\\
1020290157	&	2018-05-26T03:25:40	&	1387	&	27	\\
1020290158	&	2018-05-27T01:03:20	&	3954	&	28	\\
1020290159	&	2018-05-28T00:13:00	&	3183	&	29	\\
1020290160	&	2018-05-29T00:55:00	&	1001	&	29	\\
1020290161	&	2018-06-01T01:25:20	&	1287	&	29	\\
1020290162	&	2018-06-02T02:13:20	&	2053	&	29	\\
1020290163	&	2018-06-03T13:47:00	&	1468	&	30	\\
1020290164	&	2018-06-04T02:07:35	&	796	&	30	\\
1020290165	&	2018-06-05T02:49:35	&	1287	&	30	\\
1020290166	&	2018-06-06T20:24:43	&	1681	&	30	\\
1020290167	&	2018-06-07T05:53:13	&	426	&	30	\\
1020290168	&	2018-06-08T01:57:14	&	4238	&	30	\\
1020290169	&	2018-06-09T02:37:33	&	1767	&	31	\\
1020290170	&	2018-06-10T04:58:19	&	2928	&	31	\\
1020290171	&	2018-06-13T06:56:00	&	1941	&	31	\\
1020290172	&	2018-06-13T23:55:20	&	2161	&	31	\\
1020290173	&	2018-06-15T16:04:20	&	620	&	32	\\
1020290174	&	2018-06-16T04:25:40	&	5708	&	32	\\
1020290175	&	2018-06-17T02:02:40	&	2841	&	32	\\
1020290176	&	2018-07-02T03:10:20	&	4201	&	33	\\
1020290177	&	2018-07-03T05:23:00	&	5897	&	33	\\
1020290178	&	2018-07-04T13:45:40	&	739	&	33	\\
1020290179	&	2018-07-12T17:33:00	&	346	&	34	\\
1020290180	&	2018-07-15T19:41:40	&	1429	&	34	\\
1020290181	&	2018-07-17T01:06:20	&	3690	&	34	\\
1020290182	&	2018-07-20T04:45:40	&	895	&	35	\\
1020290183	&	2018-07-23T15:52:00	&	436	&	35	\\
1020290184	&	2018-07-24T07:19:00	&	495	&	35	\\
1020290185	&	2018-07-26T17:58:38	&	1731	&	35	\\
1020290186	&	2018-07-27T01:41:38	&	1681	&	35	\\
1020290187	&	2018-07-28T08:33:00	&	824	&	36	\\
1020290188	&	2018-07-28T23:55:40	&	2910	&	36	\\
1020290189	&	2018-08-01T11:17:42	&	1751	&	36	\\
1020290190	&	2018-08-03T15:53:00	&	580	&	37	\\
1020290191	&	2018-08-05T09:56:20	&	1016	&	37	\\
1020290192	&	2018-08-06T07:34:00	&	1780	&	37	\\
1020290193	&	2018-08-07T21:47:56	&	1328	&	37	\\
1020290194	&	2018-08-08T00:52:56	&	7480	&	37	\\
1020290195	&	2018-08-09T07:48:59	&	956	&	38	\\
1020290196	&	2018-08-10T20:53:33	&	1324	&	38	\\
1020290197	&	2018-08-10T23:58:52	&	3082	&	38	\\
1020290198	&	2018-08-12T00:38:40	&	6129	&	38	\\
1020290199	&	2018-08-13T01:22:53	&	6696	&	39	\\
1020290201	&	2018-08-14T23:34:19	&	610	&	39	\\
1020290202	&	2018-08-15T01:06:59	&	2480	&	39	\\
1020290203	&	2018-08-16T14:07:40	&	2469	&	39	\\
1020290204	&	2018-08-17T00:57:59	&	9688	&	40	\\
1020290205	&	2018-08-18T04:50:59	&	9448	&	41	\\
1020290206	&	2018-08-19T01:00:01	&	10341	&	41	\\
1020290207	&	2018-08-23T03:44:00	&	354	&	42	\\
1020290208	&	2018-08-24T00:04:59	&	2252	&	42	\\
1020290209	&	2018-08-25T00:33:00	&	4883	&	42	\\
1020290210	&	2018-08-26T01:28:20	&	1832	&	42	\\
1020290211	&	2018-08-29T07:51:20	&	3708	&	43	\\
1020290212	&	2018-09-03T08:33:20	&	1084	&	43	\\
1020290213	&	2018-09-04T19:59:40	&	1544	&	43	\\
1020290214	&	2018-10-02T00:32:59	&	3882	&	44	\\
1020290215	&	2018-10-03T01:16:16	&	6154	&	44 \\	
\enddata
\end{deluxetable}

\begin{deluxetable}{llrc}
\renewcommand\thetable{A2}
\tabletypesize{\footnotesize}
\tablecaption{Log of \textit{NICER} observations of PSR J1849$-$0001}
\label{tab:j1849_obs}
\tablecolumns{4}
\tablewidth{0pc}
\tablehead{
\colhead{Observation}   &
\colhead{Start time} &
\colhead{Exposure} & 
\colhead{TOA}\\
\colhead{ID}   &
\colhead{(UTC)} &
\colhead{(s)} & 
\colhead{Number}
}
\startdata
1020660101	&	2018-02-13T23:49:40	&	187	&	1	\\
1020660102	&	2018-02-14T01:22:20	&	504	&	1	\\
1020660103	&	2018-02-15T03:38:40	&	978	&	1	\\
1020660104	&	2018-02-16T02:45:20	&	1333	&	1	\\
1020660105	&	2018-02-17T00:49:28	&	1486	&	1	\\
1020660106	&	2018-02-19T06:35:00	&	896	&	1	\\
1020660107	&	2018-02-20T10:30:17	&	980	&	1	\\
1020660108	&	2018-02-21T05:09:40	&	498	&	1	\\
1020660109	&	2018-03-22T04:11:20	&	188	&	\nodata	\\
1020660110	&	2018-03-24T08:34:58	&	677	&	\nodata	\\
1020660113	&	2018-04-21T08:17:58	&	1176	&	2	\\
1020660115	&	2018-04-26T01:01:20	&	319	&	2	\\
1020660116	&	2018-05-01T21:53:40	&	351	&	3	\\
1020660117	&	2018-05-02T01:01:40	&	746	&	3	\\
1020660118	&	2018-05-03T00:08:20	&	2481	&	3	\\
1020660119	&	2018-05-04T00:50:00	&	5489	&	3	\\
1020660120	&	2018-05-05T09:16:00	&	1442	&	3	\\
1020660121	&	2018-05-06T00:43:40	&	34	&	3	\\
1020660122	&	2018-05-22T18:12:20	&	866	&	4	\\
1020660123	&	2018-05-23T01:55:20	&	3985	&	4	\\
1020660124	&	2018-05-24T16:45:20	&	152	&	4	\\
1020660126	&	2018-05-29T22:54:49	&	545	&	4	\\
1020660127	&	2018-05-30T00:27:29	&	2678	&	4	\\
1020660128	&	2018-06-01T00:19:30	&	2250	&	5	\\
1020660129	&	2018-06-02T04:36:00	&	210	&	5	\\
1020660130	&	2018-06-06T22:54:00	&	558	&	5	\\
1020660131	&	2018-06-08T05:38:40	&	145	&	5	\\
1020660132	&	2018-06-12T01:57:16	&	1898	&	6	\\
1020660133	&	2018-06-15T18:00:37	&	1387	&	6	\\
1020660134	&	2018-06-16T00:20:40	&	1406	&	6	\\
1020660135	&	2018-06-17T01:03:40	&	4147	&	6	\\
1020660136	&	2018-06-18T01:34:40	&	3007	&	6	\\
1020660137	&	2018-06-19T00:44:20	&	1533	&	6	\\
1020660138	&	2018-06-24T09:38:00	&	252	&	7	\\
1020660139	&	2018-06-25T05:43:20	&	595	&	7	\\
1020660140	&	2018-06-26T20:23:20	&	105	&	7	\\
1020660141	&	2018-06-27T04:04:20	&	1691	&	7	\\
1020660142	&	2018-06-28T07:54:00	&	262	&	7	\\
1020660143	&	2018-07-02T09:14:40	&	147	&	7	\\
1020660144	&	2018-07-03T08:23:20	&	423	&	7	\\
1020660145	&	2018-07-10T03:50:40	&	72	&	\nodata	\\
1020660146	&	2018-07-12T03:42:52	&	1933	&	8	\\
1020660147	&	2018-07-13T13:54:00	&	371	&	8	\\
1020660148	&	2018-07-14T03:28:41	&	1397	&	8	\\
1020660149	&	2018-07-15T02:49:40	&	832	&	8	\\
1020660150	&	2018-07-16T03:22:20	&	2096	&	8	\\
1020660151	&	2018-07-19T16:38:29	&	994	&	8	\\
1020660153	&	2018-07-22T18:21:40	&	984	&	9	\\
1020660154	&	2018-07-24T01:22:21	&	5167	&	9	\\
1020660155	&	2018-08-21T15:02:35	&	2142	&	10	\\
1020660156	&	2018-08-25T22:35:57	&	985	&	10	\\
1020660157	&	2018-08-26T00:08:37	&	288	&	10	\\
1020660158	&	2018-09-26T20:50:23	&	748	&	11	\\
1020660159	&	2018-09-26T23:55:43	&	1057	&	11	\\
1020660160	&	2018-09-28T11:28:37	&	587	&	11	\\
1020660161	&	2018-09-29T02:55:39	&	79	&	11	\\
\enddata
\end{deluxetable}

\end{document}